\documentclass[12pt]{article}
\usepackage{graphicx}
\newcommand\pubnumber{ }
\newcommand\pubdate{\today}

\def\napoli{DESY, Hamburg, Germany}
\def\support{\footnote{E-mail: {\it jan.olsson@desy.de}}}

\def\Title#1{\begin{center} {\Large #1 } \end{center}}
\def\Author#1{\begin{center}{ \sc #1} \end{center}}
\def\Address#1{\begin{center}{ \it #1} \end{center}}

\newcommand\pubblock{\rightline{\begin{tabular}{l} \pubnumber\\
          \pubdate  \end{tabular}}}
\newenvironment{Abstract}{\begin{quotation}  }{\end{quotation}}
\newenvironment{Presented}{\begin{quotation} \begin{center} 
             \end{center}\bigskip 
      \begin{center}\begin{large}}{\end{large}\end{center} \end{quotation}}
\def\Acknowledgements{\bigskip  \bigskip \begin{center} \begin{large}
             \bf ACKNOWLEDGEMENTS \end{large}\end{center}}
\def\gsim{\,\lower.25ex\hbox{$\scriptstyle\sim$}\kern-1.30ex%
\raise 0.55ex\hbox{$\scriptstyle >$}\,}
\def\lsim{\,\lower.25ex\hbox{$\scriptstyle\sim$}\kern-1.30ex%
\raise 0.55ex\hbox{$\scriptstyle <$}\,}
\def\bc{\begin{center}}
\def\ec{\end{center}}
\def\bi{\begin{itemize}}
\def\ei{\end{itemize}}
\def\bt{\begin{tabular}}
\def\et{\end{tabular}}
\def\beq{\begin{equation}}
\def\eeq{\end{equation}}

\def\NPA{Nucl. Phys.   {\bf A}}
\def\NPB{Nucl. Phys.   {\bf B}}
\def\PLB{Phys. Lett.   {\bf B}}
\def\PRL{Phys. Rev. Lett. }

\def\PRD{Phys. Rev.   {\bf D}}

\def\ZPA{Z. Phys.     {\bf A}}
\def\ZPC{Z. Phys.     {\bf C}}

\def\gprho{\gamma p \to \rho^0 n \pi^+}

\def\Wgp{W_{\gamma p}}

\def\Wgpi{W_{\gamma\pi}}

\def\sgp{\sigma_{\gamma p}}
\def\sgpi{\sigma_{\gamma\pi}}

\def\fluxpi{f_{\pi/p}}
\def\fluxpia{\fluxpi(x_L,t)}
\def\fluxg{f_{\gamma/e}}

\def\ptr{p_{T,\rho}}
\def\ptn{p_{T,n}}

\def\d2sxp{{\rm d^2}\sgp/{\rm d}x_L{\rm d}\ptn^2}
\def\beq{\begin{equation}}
\def\eeq#1{\label{#1}\end{equation}}
\def\eeqn{\end{equation}}

\def\beqa{\begin{eqnarray}}
\def\eeqa#1{\label{#1}\end{eqnarray}}
\def\eeqan{\end{eqnarray}}

\let\bar=\overbar

\def\Dslash{\not{\hbox{\kern-4pt $D$}}}
\def\dslash{\not{\hbox{\kern-2pt $\del$}}}

\def\msb{{\bar{\ssstyle M \kern -1pt S}}}

\begin{document}
\begin{titlepage}
\pubblock

\vfill
\Title{Exclusive $\rho^0$ Meson Photoproduction 
                with a Leading Neutron at HERA}
\vfill
\Author{Jan Olsson (on behalf of the H1 Collaboration)\support}
\Address{\napoli}
\vfill
\begin{Abstract}
    A first measurement is presented of exclusive photoproduction
    of $\rho^0$ mesons associated with leading neutrons at HERA.
    The data were taken with the H1 detector in the years $2006$ and $2007$
    at a centre-of-mass energy of $\sqrt{s}=319$ GeV 
    and correspond to an integrated luminosity of $1.16$ pb$^{-1}$.
    The $\rho^0$ mesons with transverse momenta $p_T<1$ GeV
    are reconstructed from their decays to charged pions, 
    while leading neutrons carrying a large fraction
    of the incoming proton momentum, $x_L>0.35$, are detected 
    in the Forward Neutron Calorimeter.
    The phase space of the measurement is defined by the photon virtuality
    $Q^2 < 2$ GeV$^2$, the total energy of the photon-proton system
    $20 < \Wgp < 100$ GeV and the polar angle of the leading neutron
    $\theta_n < 0.75$ mrad.
    The cross section of the reaction $\gamma p \to \rho^0 n \pi^+$
    is measured as a function of several variables.
    The data are interpreted in terms of a double peripheral process,
    involving pion exchange at the proton vertex followed by elastic
    photoproduction of a $\rho^0$ meson on the virtual pion. 
    In the framework of one-pion-exchange dominance 
    the elastic cross section of photon-pion  scattering,
    $\sigma^{\rm el}(\gamma\pi^+ \to \rho^0\pi^+)$, is extracted. 
    The value of this cross section indicates significant
    absorptive corrections for the exclusive reaction $\gprho$.
\end{Abstract}
\vfill
\begin{Presented}
Presented at EDS Blois 2017, Prague, \\ Czech Republic, June 26-30, 2017
\end{Presented}
\vfill
\end{titlepage}
\def\thefootnote{\fnsymbol{footnote}}
\setcounter{footnote}{0}

\section{Introduction}
At the $ep$ collider HERA the study of forward production
of protons and neutrons has long been a subject of
interest. The term ``forward'' here implies that the produced
particles have small polar angles with respect to the proton beam direction
and carry a large fraction of the incoming proton longitudinal momentum.
The measurement of these leading baryons, together with their associated
final states, provide an important input to theoretical models for describing
the strong interaction in non-perturbative regimes.
\par\noindent
In the present report this long tradition is continued with a study of the
exclusive photoproduction of $\rho$ mesons. This analysis has recently been
published \cite{h1rhoneutron}; for more details than can be given here 
of the experimental apparatus, data and 
kinematic variables as well as of the analysis method, please see this paper.
\par\noindent
The diagram of the reaction 
\begin{equation}
ep \to e \rho n \pi^+
\end{equation}
is depicted in Fig.~1a. The virtual
photon from the electron forms a virtual $\rho$ meson, which scatters 
elastically on a
pion in the ``proton cloud'', the final state contains a $\rho^{\circ}$ meson  
together with the scattered $\pi^+$ and the leading neutron. 
It is clear that a measurement of this reaction offers the possibility of 
measuring the exclusive elastic photoproduction of rho mesons off a pion
target, an unstable particle. Such an experiment was suggested already in 1959
by G.F.~Chew and P.E.~Low.
\begin{figure}[htb]
\centering
\includegraphics[height=1.39in]{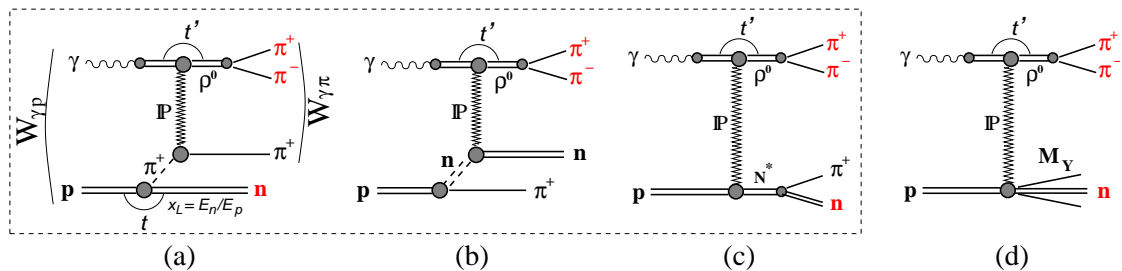}
\caption{Generic diagrams for processes contributing to exclusive
         photoproduction of $\rho^0$ mesons associated with leading neutrons 
         at HERA.
         The signal corresponds to the Drell-Hiida-Deck model graphs 
         for the pion exchange (a), neutron exchange (b) and 
         direct pole (c).  
         Diffractive scattering in which a neutron may be produced as 
         a part of the
         proton dissociation system, $M_Y$, contributes as background (d).
         The $N^*$ in (c) denotes both resonant (via $N^+$) and possible 
         non-resonant $n+\pi^+$ production.}
\label{fig:F1}
\end{figure}
\par\noindent
In the Regge formalism, 
this $2 \to 3$ process, commonly known as a 
Double Peripheral Process (DPP)\cite{DPP}, is
described by the exchange of two Regge trajectories. 
Thus, the proton dissociates into a neutron
and a Regge $\pi^+$, and the ($n,\pi^+$) system scatters elastically 
off the virtual $\rho$ meson, with the exchange of the Regge trajectory 
with vacuum quantum numbers, the ``Pomeron''.
\par\noindent
Double peripheral processes 
were in the past extensively studied at lower energies, 
in nucleon-nucleon and meson-nucleon scattering. The theoretical framework
was the Drell-Hiida-Deck (DHD) model\cite{Deck} and its generalisation. 
According to this model, also the diagrams in Figs.~1b and 1c have to be 
considered, as well as the interference between the three diagrams.
The diagrams~1b and 1c, describing neutron exchange and direct pole 
respectively, contribute in 
the cross section with similar magnitude, however with 
opposite sign. In the phase space region considered in this analysis, with
low transverse momentum of the produced leading neutron, 
the diagrams~1b and 1c largely cancel\cite{Tsarev}  and the 
One Pion Exchange\cite{OPE} diagram of Fig.~1a dominates. 
Of the possible isovector
exchanges ($\pi, \rho, a_2$), the pion dominates and the OPE approximation
is expected to give a good description of the process.
\section{Analysis}
\par\noindent
The experimental signature of reaction (1) is the scattered neutron, as well
as the $\pi^+\pi^-$ decay products of the rho meson. The scattered $\pi^+$ 
is not
detected, since it is produced at low transverse momentum and escapes in the
beam tube. The scattered electron is also not detected (no tag experiment), 
which limits
the $Q^2$ range to $Q^2<2 {\rm GeV}^2$ (photoproduction regime at HERA, with 
$\langle Q^2 \rangle = 0.04$ GeV$^2)$. 
\par\noindent
Background to the signal reaction (1) is given by the low mass dissociation 
of the proton (Fig.1d), as well as by reflections from the exclusive 
production of other vector mesons, in particular $\omega$, 
$\phi$ and $\rho^{\prime}(1450-1700)$.
\begin{figure}[htb]
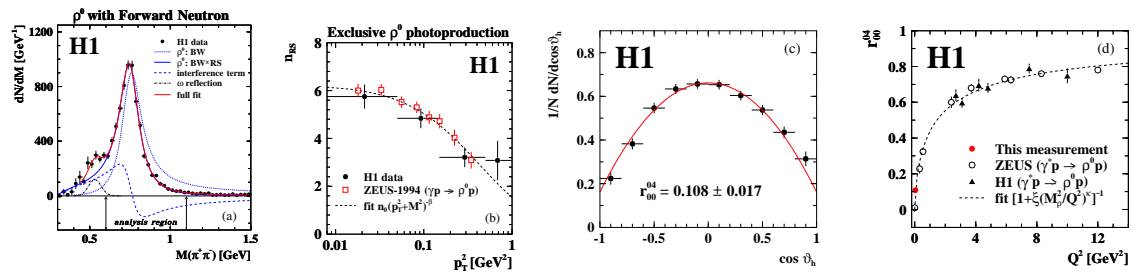

\centering
\includegraphics[height=1.39in]{d15-120f3a.eps}
\includegraphics[height=1.39in]{d15-120f3b.eps}
\includegraphics[height=1.39in]{d15-120f3c.eps}
\includegraphics[height=1.39in]{d15-120f3d.eps}
\caption{(a) Mass distribution of the $\pi^+\pi^-$ system. 
         (b) Ross-Stodolsky skewing parameter, $n_{RS}$, as a function
         of $p_T^2$ of the $\pi^+\pi^-$ system. 
         (c) Decay angular distribution of the $\pi^+$ in the helicity frame.
         (d) Spin-density matrix element, $r_{00}^{04}$, as a function
         of $Q^2$ for diffractive $\rho^0$ photo- and
         electro-production.}
\label{fig:F2}
\end{figure}
\par\noindent
For the MC simulation of the OPE signal process (Fig~1a) the program
POMPYT\cite{Pompyt}
is used. The background (Fig.~1d) is simulated using the program
DIFFVM\cite{DiffVM}, 
which is also used for estimating possible background due to reflections 
from other exclusive vector meson production.
\par\noindent
The data in this analysis were obtained using a special low bias trigger, 
based on the H1 Fast Track Trigger in connection with a neutron signal in
the H1 FNC (Forward Neutron Calorimeter). The trigger was operated in downscale
mode during the last two years of HERA operation, 2006-2007, and the data
sample corresponds to an integrated luminosity of 1.16~pb$^{-1}$.
%
%%%%%%%%% Table 1
%
\begin{table}[hbt]
        \centering
        \renewcommand{\arraystretch}{1.30}
\bt{|l r|c|c|}
 \hline %--------------------------------------------------------------------------------
     \multicolumn{2}{|l|}{Event selection} &  {Analysis PS}   &    {Measurement PS}           \\
 \hline %--------------------------------------------------------------------------------
    {No} $e'$ {in the detector}  &  &   $Q^2 < 2$ {GeV}$^2$    &    $Q^2 = 0$ {GeV}$^2$ \          \\
               &  & $\langle Q^2 \rangle = 0.04$ {GeV}$^2$   &                             \\
 \hline %--------------------------------------------------------------------------------
    $2$ {tracks, net charge} $=0$ & &     &                             \\
    ~ $p_T\!>\!0.2$ {GeV},~ $20^o\!<\!\theta\!<\!160^o$,     &
                 &  $20<\Wgp<100$ {GeV}  &  $20<\Wgp<100$ {GeV}     \\
               &  & $\langle W_{\gamma p} \rangle = 45$ {GeV}   &       \\
    ~ {from} $|z_{\rm vx}|<30$ {cm} & & $\ptr<1.0$ {GeV}        
 &  $-t^{\prime}<1.0$ {GeV}$^2$  \\
    $0.3<M_{\pi\pi}<1.5$ {GeV}    & & $0.6<M_{\pi\pi}<1.1$ {GeV} 
 & $2m_{\pi}<M_{\rho}< M_{\rho}\!+\!5\Gamma_{\rho}$ \\
 \hline %--------------------------------------------------------------------------------
         {LRG requirement}       & & $\sim 637,000$ {events}   &                             \\
 \hline %--------------------------------------------------------------------------------
         {Neutron requirements}       & &   &                            \\
    $E_n>120$ {GeV}              & & $x_L > 0.2$      & $0.35<x_L<0.95$  \\
    $\theta_n<0.75$ {mrad}       & & $\theta_n<0.75$ {mrad}    & $\ptn<x_L\cdot 0.69$ {GeV}    \\
%    $x_{FNC}\!<\!2.5$cm,
%    $y_{FNC}\!<\!7.5$cm        &                          &               \\
 \hline %--------------------------------------------------------------------------------
    $\sim 7000$ {events}       &  & $\sim 6100$ {events}       & $\sim 5770$ {events}         \\
 \hline %--------------------------------------------------------------------------------
  {OPE dominated range} & & $\ptn\!<\!0.2$ {GeV}     &  $\sim 3600$ {events} \\
   \hline %--------------------------------------------------------------------------------
\et
%  \vspace{\baselineskip} 
        \caption{Event selection criteria and the definition 
                 of the kinematic phase space (PS) of the measurements.
                 The measured cross sections are determined at $Q^2=0$
                 using an effective flux based on the VDM.}
%
%   possibly add a reference here ???
%
  \label{tab:selection}
\end{table}
\par\noindent
After all experimental cuts (see Table 1), $\sim 5770$ events 
with two oppositely
charged tracks and a leading neutron form the final data sample. The pi+pi-
mass distribution is shown in Fig.~2a, together with curves of the 
Breit-Wigner fit, using the Ross-Stodolsky skewing model\cite{RS}. 
The Ross-Stodolsky
parameter $n_{RS}$ is shown in Fig.~2b as a function of the transverse momentum
squared of the $\pi^+\pi^-$ system. The measured value agrees with previous
measurements of the ZEUS collaboration.
Further corroboration of the $\rho$ signal is given by Fig.~2c, showing the
$\rho$ helicity frame decay angle distribution, and Fig.~2d, showing the 
extracted
spin-density matrix element $r^{04}_{00}$. The present 
measurement is compared with previous measurements at HERA, 
at various values of $Q^2$.
\par\noindent
The general conclusion from Fig.~2 is that the $\pi^+\pi^-$ data sample 
shows the characteristics  of $\rho$ photoproduction.
The fraction of diffractive background (Fig.~1d)
was determined to be $0.34 \pm 0.05$.
\par\noindent
The cross section of reaction (1), which is 
directly measured in this experiment, is related to the photoproduction
cross section:
\begin{equation}
{\rm d}^2\sigma_{ep}/{\rm d}y {\rm d}Q^2 = \fluxg(y,Q^2) \sgp(\Wgp(y)).
\label{eq:ep2gp}
\end{equation}  
where the virtual photon flux is taken from the VDM\cite{Sakurai}. 
The $\gamma p$ cross section can be written as a convolution of the 
$\gamma\pi^+$ cross section and the pion flux $f_{\pi/p}$, using the
OPE approximation:
\begin{equation}
{\rm d}^2\sgp(\Wgp,x_L,t)/{\rm d}x_L{\rm d}t = \fluxpia\,\sgpi(\Wgpi). 
\label{eq:gp2gpi}
\end{equation}  
\par\noindent
$f_{\pi/p}$ describes the $n,\pi$ splitting of the proton.
From the many existing models of the pion flux\cite{pionflux}, 
the one of H.~Holtmann et~al. 
is used for the central value of $\sigma_{\gamma\pi^+}$. 
The use of other models 
indicates a 30\% systematic error due to the uncertainty in the pion flux.
\section{Results}
\par\noindent
Cross sections are measured in the analysis phase space, i.e. 
$Q^2 < 2$ GeV$^2$, $20 < \Wgp < 100$ GeV and $\theta_n < 0.75$ mrad.
The $\gamma p$ cross section, integrated over 
$0.35<x_L<0.95$ and $p_T^{\rho}<1$ GeV, 
is determined for two regimes of the leading neutron transverse momentum,
resulting in the following average values over the $W_{\gamma p}$ range:
\begin{equation}
  \sigma (\gamma p \to \rho^0 n \pi^+ ) = (310 \pm 6_{\rm stat} \pm 45_{\rm sys})~ {\rm nb}
         \hspace*{0.8cm} {\rm for} \hspace*{0.3cm} \ptn<x_L \cdot 0.69 {\rm ~GeV}
  \label{eq:sgp1}
\end{equation}
\begin{equation} 
 \sigma (\gamma p \to \rho^0 n \pi^+ ) = (130 \pm 3_{\rm stat} \pm 19_{\rm sys})~ {\rm nb} 
         \hspace*{0.8cm} {\rm for} \hspace*{1.2cm} \ptn<0.2 {\rm ~GeV}.  
  \label{eq:sgp2}
\end{equation}

\par\noindent
The second cross section, obtained for the stricter cut $\ptn<0.2 {\rm ~GeV}$
(OPE dominated regime), is used to extract the $\gamma \pi^+$ cross section:
\begin{equation}
 \sigma (\gamma \pi^+\to \rho^0\pi^+) = 
 (2.33\pm0.34 (\rm exp) ^{+0.47}_{-0.40} (\rm model))~\mu\rm b \hspace*{0.5cm}
  {\rm with} \hspace*{0.3cm} \langle \Wgpi \rangle \simeq 24~{\rm GeV}
 \label{eq:sgp3}
\end{equation}
\par\noindent
The differential cross section $d\sigma_{\gamma p}/dx_L$   
is shown in Fig.~3\footnote{An overall normalisation error of 4.4\% is not
included in this and the following figures.}. 
Several models of the pion flux are compared to the data.
As seen, two of the six models (FMS, NSSS) agree badly with the data and can
be excluded, while the other show reasonable agreement.   
\par\noindent
Sensitivity to the pion flux models is also visible in the $t$ 
(or $p^2_{t,n}$)
dependence of the leading neutron. Fig.~4 (left) 
shows the double differential cross
section $d\sigma_{\gamma p}/dx_L dp^2_{t,n}$. The bins are chosen such that 
the result is not influenced by the cut in neutron polar angle. The $x_L$ 
dependence of the slope $b_n$, obtained from the fits of the 
single exponential function
$e^{-b_n(x_L)\ptn^2}$ to the data, is also shown in Fig.~4 (right). None of the
flux models fits the data. A possible explanation of this would be 
absorptive corrections which modify the $t$ dependence of the amplitude 
such that the effective slope increases for large values of $x_L$, compared
to the pure OPE model without absorption. The same slope dependence was 
observed by the ZEUS collaboration\cite{Chekanov:2007} in their study of 
inclusive DIS data with a leading neutron.
\begin{figure}[htb]
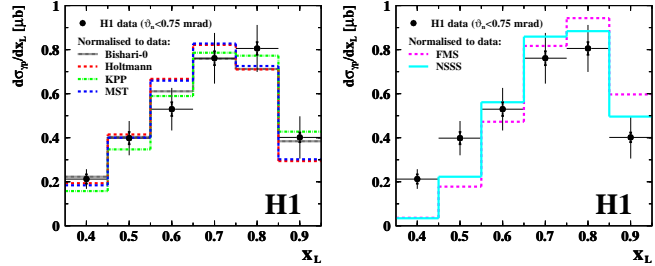

\centering
\includegraphics[height=1.5in]{d15-120f6b.eps}
\includegraphics[height=1.5in]{d15-120f6a.eps}
\caption{Differential cross section d$\sgp/{\rm d}x_L$ 
       in the range $20<\Wgp<100$ GeV compared
       to the predictions based on different versions of the pion flux models.
       Left: disfavoured versions of the pion flux models. 
       Right: pion flux models compatible with the data. 
       All predictions are normalised to the data. }
\label{fig:F3}
\end{figure}
\begin{figure}[htb]
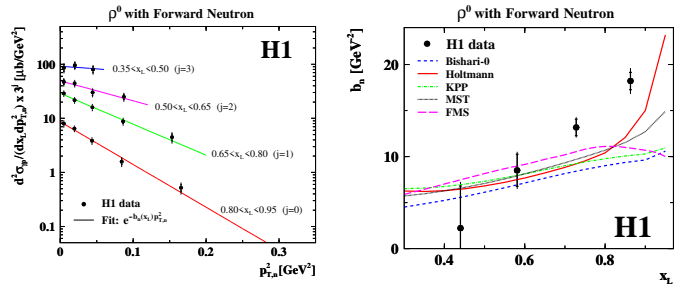

\centering
\includegraphics[height=1.5in]{d15-120f7.eps}
\includegraphics[height=1.5in]{d15-120f8.eps}
\caption{Left: Double differential cross section $\d2sxp$ of neutrons
         in the range $20<\Wgp<100$ GeV 
         fitted with single exponential functions.
         Right: The exponential slopes 
         fitted through the $p_T^2$ dependence of the leading neutrons
         as a function of $x_L$.
         The expectations of  
         several parametrisations of the pion flux within the
         OPE model are compared to the data.     }
\label{fig:F4}
\end{figure}
\par\noindent
The measured $W_{\gamma p}$ dependence of $\sigma(\gamma p \to \rho^0 n \pi^+)$
is shown in Fig.~5 (left), 
together with the prediction from the POMPYT MC simulation.
The measured cross section falls with increasing $W_{\gamma p}$, in contrast
to the POMPYT simulation, which is slowly increasing with energy, 
due to the pomeron exchange. 
Fitting the cross section with the Regge motivated function
$W^\delta_{\gamma p}$, one obtains 
$\delta = -0.26 \pm 0.06_{stat} \pm 0.07_{sys}$.
\begin{figure}[htb]
\centering
\includegraphics[height=1.5in]{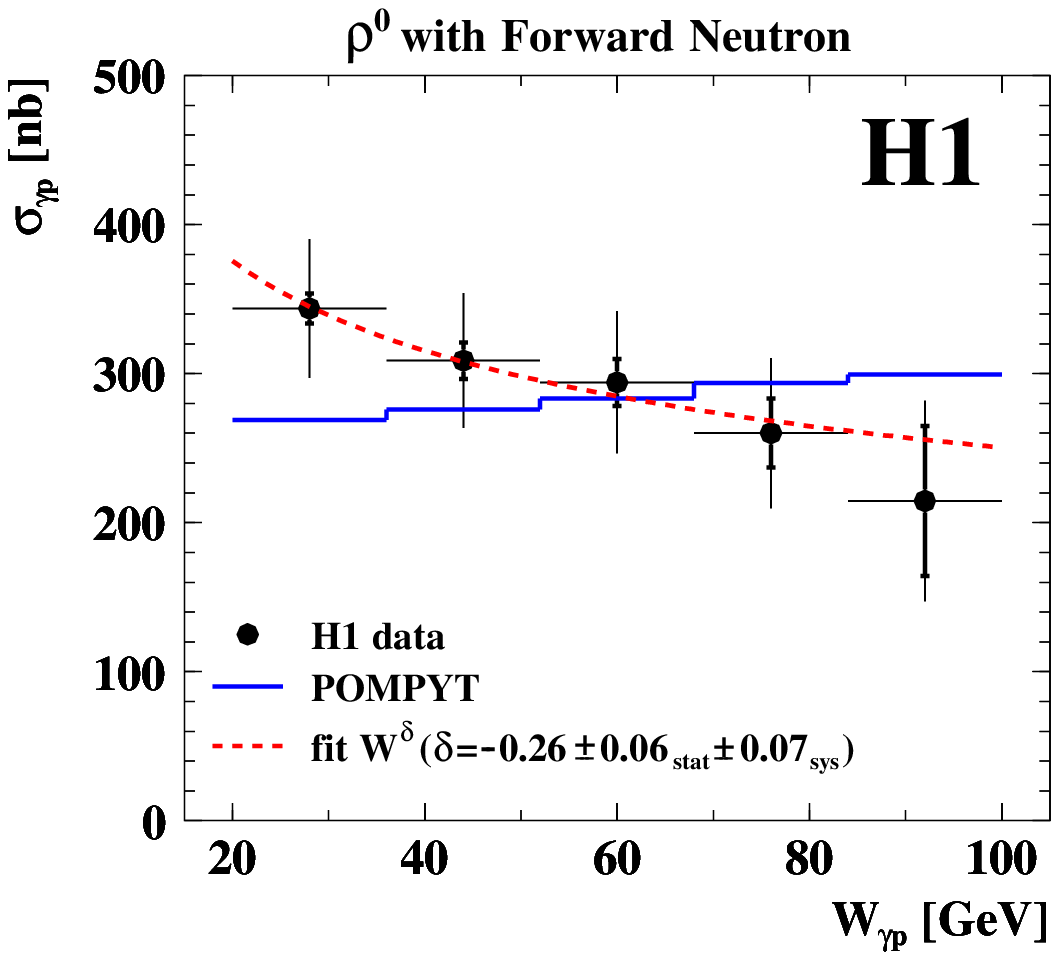}
\includegraphics[height=1.5in]{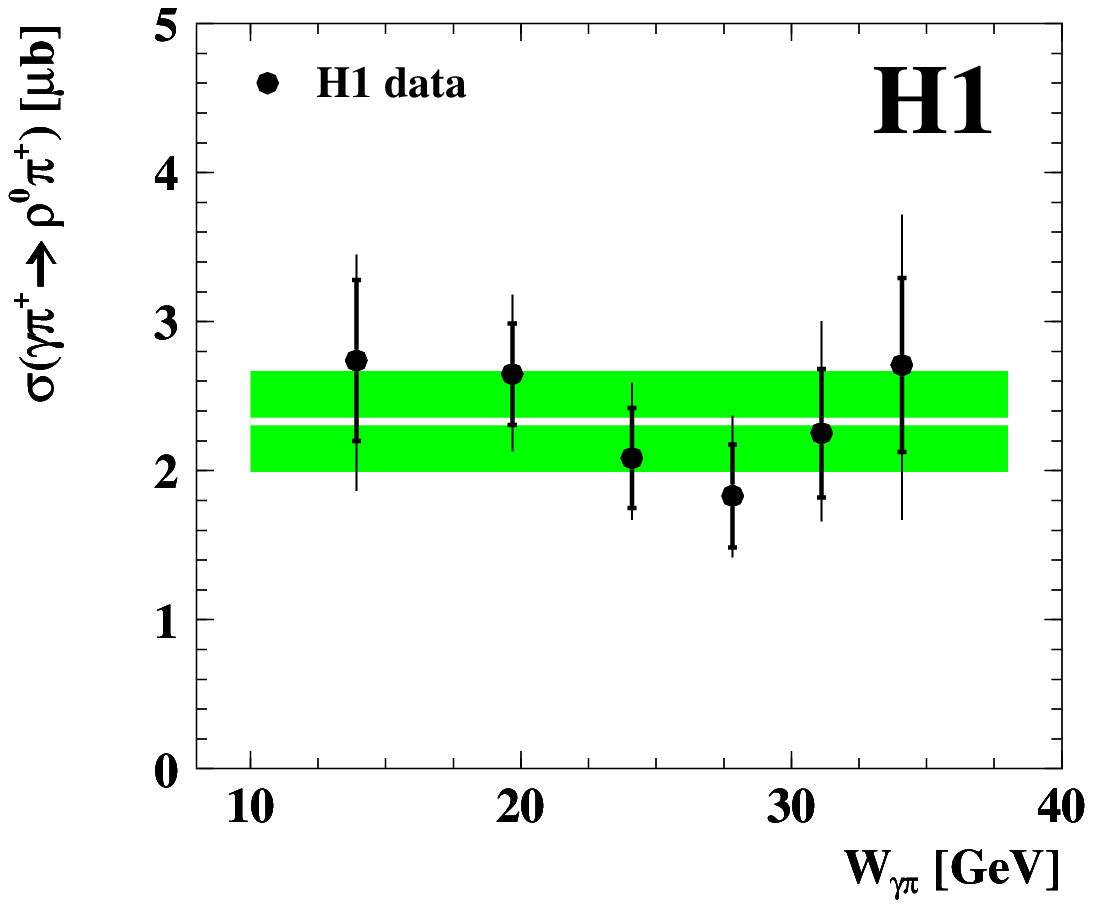}
\includegraphics[height=1.5in]{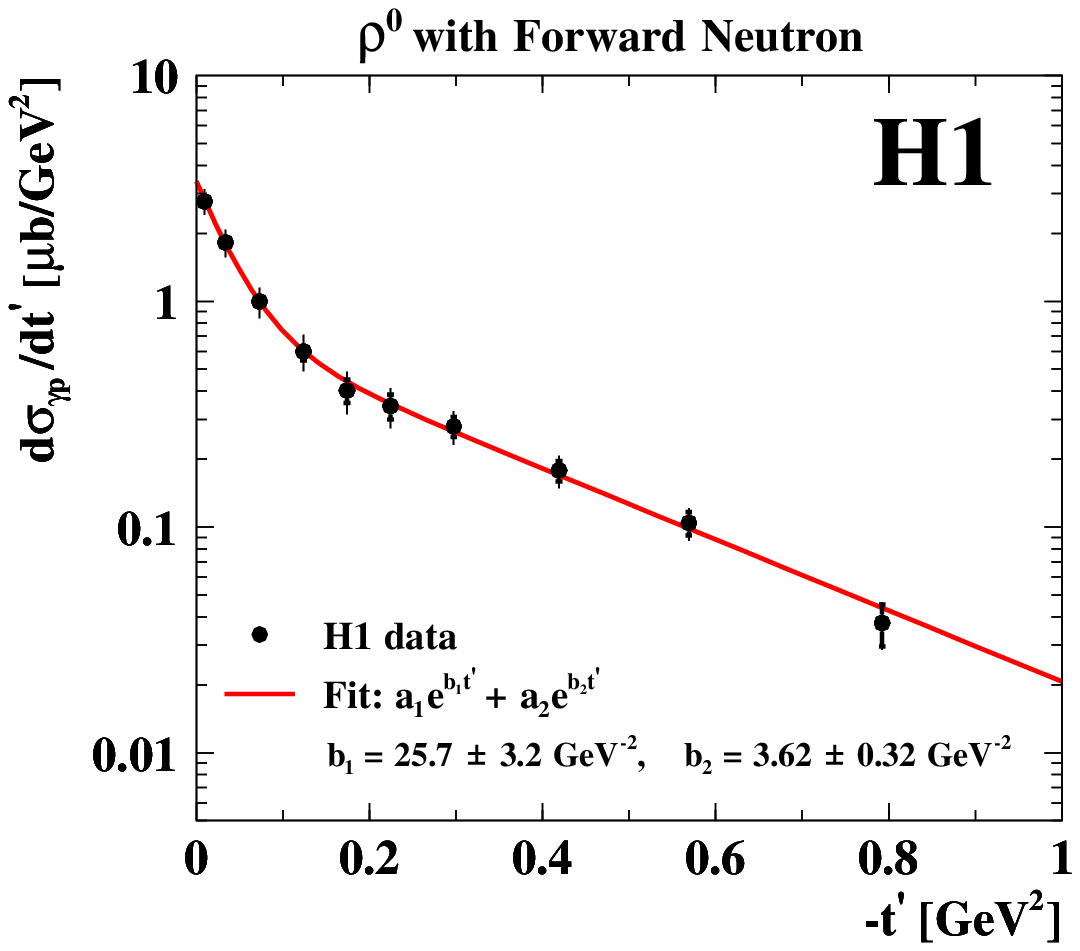}
\caption{Left: Cross section of the reaction $\gprho$ as
         a function of $\Wgp$ compared to the prediction 
         from the POMPYT MC program, which is normalised to the data.
         The dashed curve shows the fit of the function
         $\sigma \propto W^{\delta}$ to the data,
         with $\delta = -0.26 \pm 0.06_{\rm stat}\pm 0.07_{\rm sys}$.
         Middle: The elastic cross section
         $\sgpi^{\rm el} \equiv \sigma ({\gamma\pi^+} \to \rho^0\pi^+)$,
         extracted in the one-pion-exchange approximation,
         as a function of the photon-pion energy, $\Wgpi$.
         The dark shaded band represents the average value for the full $\Wgpi$
         range.
         Right: Differential cross section 
         $d\sigma_{\gamma p}/{\rm d}t^{\prime}$,
         fitted with the sum of two exponential functions. }
\label{fig:F5}
\end{figure}
\par\noindent
The $W_{\gamma \pi^+}$ dependence of $\sigma (\gamma \pi^+\to \rho^0\pi^+)$,
displayed in Fig.~5 (middle), does not show a clear tendency.  
\par\noindent
The value of $\sigma (\gamma \pi^+\to \rho^0\pi^+)$, obtained at the 
average value $\langle \Wgpi \rangle \simeq 24$~GeV, 
can be compared to the
corresponding cross section for $\gamma p \to \rho^0 p$, obtained at
$\langle \Wgp \rangle \simeq 24$~GeV\footnote{This value is used to 
interpolate between the fixed target measurements and the HERA measurements.}. 
The ratio of the measured elastic cross sections is 
$r_{\rm el} = \sigma_{\rm el}^{\gamma\pi}/\sigma_{\rm el}^{\gamma p} = 0.25 \pm 0.06$.  
While the additive quark model would
predict a value 2/3 for this ratio, more sophisticated considerations (using
the optical theorem and eikonal approach, involving both $\gamma p$, $p p$ and
$\pi p$ cross sections) lead to the expectation $0.57 \pm 0.03$. 
An explanation for the measured reduced ratio is the 
rescattering (absorptive corrections\cite{abs_corr}), involved
in the leading neutron production. For the reaction (1) studied here, an
absorption factor $0.44 \pm 0.11$ is obtained.
\par\noindent
For the total cross section at
$\langle \Wgpi \rangle \simeq 107$ GeV\footnote{Assuming a Regge dependence of 
$\sigma_{tot}^{\gamma\pi}$ and comparing with $\sigma_{tot}^{\gamma p}$ at
$\langle \Wgp \rangle \simeq 209$~GeV.} 
the ZEUS collaboration\cite{Chekanov:2002} obtained the ratio 
$r_{\rm tot} = \sigma_{\rm tot}^{\gamma\pi}/\sigma_{\rm tot}^{\gamma p} = 0.32 \pm 0.03$.  
\par\noindent
Finally, the $t^{\prime}$ distribution, i.e. the distribution of the
momentum transfer squared of the $\rho$ meson, is measured. It is shown in
Fig.~5 (right). There is a clear change in slope between the low-value and 
high-value $t^{\prime}$ ranges. The data are fitted with the sum of two
exponential functions, 
$d\sigma_{\gamma p}/dt^{\prime} = a_1e^{b_1t^{\prime}} + a_2e^{b_2t^{\prime}}$,
with the fit yielding the slope parameters 
$b_1 = (25.72 \pm 3.22_{unc} \pm 0.26_{cor})~{\rm GeV}^{-2}$ and
$b_2 = (3.62 \pm 0.30_{unc} \pm 0.10_{cor})~{\rm GeV}^{-2}$.
\par\noindent
The large value of $b_1$ indicates that most of the $\rho$ production happens
at large impact parameters, i.e. in the pion cloud extending well outside the 
classical proton radius 
($\langle r^2\rangle = 2b_1\!\cdot\!(\hbar c)^2 \simeq 2 {\rm fm}^2 \approx (1.6 R_{\rm p})^2$). In contrast, 
the small value of the second slope $b_2$ corresponds to a
target size of $\sim\!0.5$ fm.
\par\noindent
The DPP interpretation predicts a cross section dependence on the ($n,\pi^+$)
mass, resulting from interference of the 3 diagrams 
in Fig.1a-c\cite{interf_1}. Since the 
scattered $\pi^+$ is undetected in the present experiment, the invariant
($n,\pi^+$) mass and the dependence of the slope $b$ on this mass
cannot be measured and a closer investigation is not possible.
%%%%%%%%%%%%%%%%%%%%%%%%%%%%%%%%%%%%%%%%%%%%%%%%%%%%%%%%%%%%%%%%%%%%%%%%%
%%
%%   use this format to include an .eps figure into your paper
%%
% \begin{figure}[htb]
% \centering
% \includegraphics[height=1.5in]{magnet}
% \caption{Plan of the magnet used in the mesmeric studies.}
% \label{fig:youfigure}
% \end{figure}
%%%%%%%%%%%%%%%%%%%%%%%%%%%%%%%%%%%%%%%%%%%%%%%%%%%%%%%%%%%%%%%%%%%%%%%%%%%
%%%%%%%%%%%%%%%%%%%%%%%%%%%%%%%%%%%%%%%%%%%%%%%%%%%%%%%%%%%%%%%%%%%%%%%%%
%%
%%   use this format to include a LaTeX table  into your paper
%%
% \begin{table}[t]
% \begin{center}
% \begin{tabular}{ccc}  
%  ...
% \end{tabular}
% \caption{Blood cyanide levels for the two patients.}
% \label{tab:blood}
% \end{center}
% \end{table}
%%%%%%%%%%%%%%%%%%%%%%%%%%%%%%%%%%%%%%%%%%%%%%%%%%%%%%%%%%%%%%%%%%%%%%%%%%%
%
%
\vspace{-0.5cm}
\Acknowledgements 
\par\noindent
Many thanks to all colleagues in H1, in particular to Sergey Levonian and 
to the
late Iakov Vazdik, for providing the material in this report and for help
given in its preparation. Warm thanks also to the EDS 2017 team for the 
excellent organisation and the pleasant atmosphere of the conference.

\end{document}